\documentclass[letterpaper, 10 pt, conference]{ieeeconf}

\IEEEoverridecommandlockouts                           
\usepackage{graphics}
\usepackage{epsfig}
\usepackage{times}
\usepackage{amsmath}
\usepackage{amssymb}
\usepackage{hyperref} 
\usepackage{color}
\usepackage[T1]{fontenc}
\usepackage{algorithm}
\usepackage{algpseudocode}
\usepackage{multirow}
\usepackage{ctable}
\usepackage{subcaption}
\usepackage{amsfonts}
\usepackage{footnote}
\usepackage{cite}

\let\labelindent\relax

\usepackage{enumitem}

\DeclareMathOperator*{\argmin}{arg\!\min}
\renewcommand{\baselinestretch}{0.916}

\title{\bf A Comparison between Markov Chain and Koopman Operator Based Data-Driven Modeling of Dynamical Systems}

\author{Saeid Tafazzol, Nan Li, Ilya Kolmanovsky, Dimitar Filev
\thanks{S. Tafazzol and N. Li are with the Department of Aerospace Engineering, Auburn University, Auburn, AL 36849. {\tt\small szt0097,} {\tt\small nzl0058@auburn.edu}. I. Kolmanovsky is with the Department of Aerospace Engineering, University of Michigan, Ann Arbor, MI 48105. {\tt\small ilya@umich.edu}. D. Filev is with the Hagler Institute, Texas A\&M University, College Station, Texas 77843. {\tt\small dfilev@tamu.edu}.
}\vspace{-3mm}}

\begin{document}

\maketitle
\thispagestyle{empty}
\pagestyle{empty}

\begin{abstract}
Markov chain-based modeling and Koopman operator-based modeling are two popular frameworks for data-driven modeling of dynamical systems. They share notable similarities from a computational and practitioner's perspective, especially for modeling autonomous systems. The first part of this paper aims to elucidate these similarities. For modeling systems with control inputs, the models produced by the two approaches differ. The second part of this paper introduces these models and their corresponding control design methods. We illustrate the two approaches and compare them in terms of model accuracy and computational efficiency for both autonomous and controlled systems in numerical examples.
\end{abstract}

\vspace{-1mm}
\section{Introduction}

As engineered systems are increasingly more complex and cyber-physical, deriving their models using conventional first-principle approaches is becoming more difficult. Meanwhile, data-driven modeling approaches are attracting increasing attention from both researchers and practitioners. Among the variety of data-driven approaches, we are particularly interested in two approaches: Markov chain-based modeling~\cite{norris1998markov} and Koopman operator-based modeling \cite{mauroy2020koopman}. These two approaches share some common advantages that are important for their applications to modeling complex real-world systems especially for control purposes, including: 1)~the ability to model linear and nonlinear dynamics in a uniform manner, with fast ``linear-like'' dynamic flow prediction; 2) good interpretability; and 3) for controlled systems, they lead to models which facilitate the application of well-established, computationally-efficient methods for designing controls, e.g., discrete dynamic programming \cite{puterman2014markov} and linear model predictive control (MPC) \cite{rawlings2017model}.

In this paper, we first illuminate the similarities between Markov chain-based and Koopman operator-based approaches, especially for modeling autonomous systems (i.e., dynamical systems without control inputs). While the theories behind the two approaches are rich and have close connections \cite{lasota1998chaos}, we introduce the similarities mainly from a computational and practitioner's perspective. Then, for controlled systems (i.e., dynamical systems with control inputs), we show that the models produced by the two approaches differ, and different methods can be used to design controls. We illustrate and compare the two approaches for both autonomous and controlled systems through numerical examples based on the Van der Pol oscillator, which is known to be a strongly nonlinear dynamical system.

This paper is organized as follows: In Section~II, we introduce Markov chain-based modeling and Koopman operator-based modeling of autonomous dynamical systems, with an emphasis on elucidating their similarities from a computational and practitioner's perspective. In Section~III, we introduce the two approaches for modeling controlled systems and their corresponding control design methods. In~Section~IV, we illustrate and compare the two approaches through numerical examples. The paper is concluded in Section~V.

\vspace{-1mm}
\section{Markov chain and Koopman operator-based modeling of autonomous systems}

In this section, we introduce Markov chain-based and Koopman operator-based modeling of autonomous dynamical systems. We represent the system to be modeled~as:
\begin{equation}\label{equ:1}
\Sigma: x \to x^+
\end{equation}
where $x$ denotes the system state at a certain time instant, $x^+$~the system state at the corresponding next time instant, and $\to$ represents a time-based or event-based transition from~$x$ to $x^+$. We assume that the system state takes values in $\mathbb{R}^n$. The transition $\to$ may be deterministic or stochastic due to uncertainties or noises acting on the system.

\vspace{-1mm}
\subsection{Markov chain-based modeling}

A Markov chain models the dynamics of a given system as a stochastic process and describes the transition of the system state between two consecutive time instants using transition probabilities of the form:
\begin{equation}\label{equ:2}
p_{ij} = \mathbb{P}(s^+ = s_i \,|\, s = s_j)
\end{equation}
where $s$ and $s^+$ are representations of $x$ and $x^+$ in the state space of the Markov chain, called ``Markov state'' in this paper, $s_i$ and $s_j$ denote their two possible values, and $\mathbb{P}(\cdot \,|\, \cdot)$ represents a conditional probability.

Although the general Markov chain theory extends to infinite state spaces \cite{meyn2012markov}, due to computational reasons, in practical applications it is typical to consider a finite state space $\mathcal{S} = \{s_1,s_2,\dots,s_N\}$. 
Now, define a column vector $\pi \in \mathbb{R}^N$ to represent the probability distribution of the Markov state~$s$ over the state space $\mathcal{S}$ and define a matrix $P \in \mathbb{R}^{N \times N}$ that collects all transition probabilities, i.e., \small
\begin{equation}\label{equ:3}
\pi = \begin{bmatrix} \mathbb{P}(s = s_1) \\ \mathbb{P}(s = s_2) \\ \vdots \\ \mathbb{P}(s = s_N) \end{bmatrix} \quad P = \begin{bmatrix} p_{11} & p_{12} & \cdots & p_{1N} \\ p_{21} & p_{22} & \cdots & p_{2N} \\
\vdots & \vdots & \ddots & \vdots \\
p_{N1} & p_{N2} & \cdots & p_{NN} \end{bmatrix}
\end{equation} \normalsize
Then, a Markov chain model can be compactly expressed as
\begin{equation}\label{equ:4}
\pi^+ = P \pi
\end{equation}
which is a linear system of the state distribution vector $\pi$.

In order to represent the system state in $\mathbb{R}^n$ using (a combination of) finite Markov states $s_i \in \mathcal{S}$, one needs to map each point $x$ of $\mathbb{R}^n$ to a probability distribution on $\mathcal{S}$, called ``encoding.'' The encoding process can be written as
\begin{equation}\label{equ:5}
\pi = g_{\text{mc}}(x)
\end{equation}
where the entries of $g_{\text{mc}}: \mathbb{R}^n \to \mathbb{R}^N$, denoted as $g_{\text{mc},i}$, $i = 1,\dots,N$, are referred to as ``encoding functions'' or ``membership functions,'' because they quantify the degree of belonging of $x$ to each Markov state $s_i \in \mathcal{S}$.

One possible choice of encoding functions is indicator/step functions, $g_{\text{mc},i}(x) = \mathbb{I}_{X_i}(x)$, where $X_i$, $i = 1,\dots,N$, are subsets of $\mathbb{R}^n$ and form a partition of of $\mathbb{R}^n$ (i.e., $\bigcup_{i=1}^N X_i = \mathbb{R}^n$ and $X_i \cap X_j = \emptyset$ for $i \neq j$), and $\mathbb{I}_{X_i}(x) = 1$ if $x \in X_i$ and $\mathbb{I}_{X_i}(x) = 0$ if $x \notin X_i$. Such a choice corresponds to the conventional discretization strategy that partitions $\mathbb{R}^n$ into disjoint cells and maps all points in the same cell to the same Markov state. Other encoding functions may also be considered. For instance, in \cite{filev2013generalized} and \cite{li2020fuzzy}, it is shown that using Gaussian kernel functions as encoding functions can lead to a Markov chain model which provides improved prediction accuracy compared to one based on cell partition. A Gaussian kernel encoding function is given as:
\begin{equation}\label{equ:6}
g_{\text{mc},i}(x) = \frac{1}{c}\, \text{exp} \big(-\frac{1}{2} (x - \bar{x}_i)^{\top} \Sigma^{-1}_i (x - \bar{x}_i)\big)
\end{equation}
where $\bar{x}_i$ and $\Sigma_i$ are the center and the covariance of the Gaussian distribution and are design parameters, and $c$ is a normalization factor to make the output $\pi = g_{\text{mc}}(x)$ a probability distribution (i.e., $\sum_{i = 1}^N \pi_i = \sum_{i = 1}^N g_{\text{mc},i}(x) = 1$).

After the encoding functions are chosen, calibrating the Markov chain model using system trajectory data reduces to finding a transition matrix $P$ that best fits the data. This calibration process can be expressed as the solution process to the following optimization problem:
\begin{equation}\label{equ:7}
\min_{P \in \Omega}\, \mathcal{L}\left(\left\{(\pi^{k,+} - P\pi^k)\right\}_{k = 1, \dots, K} \right)
\end{equation}
where $\Omega = \{P \in \mathbb{R}^{N \times N}: P_{ij} \ge 0, \sum_{i = 1}^N P_{ij} = 1\}$ denotes the set of all probability transition matrices, $\pi^k = g_{\text{mc}}(x^k)$ and $\pi^{k,+} = g_{\text{mc}}(x^{k,+})$, $k = 1, \dots, K$, are encoded trajectory data, which are encoding function values at trajectory data pairs $(x^k, x^{k,+})$, and $\mathcal{L}(\cdot)$ is a loss function that quantifies the error of the Markov chain model. One possible choice of the loss function is
\begin{equation}\label{equ:8}    
\mathcal{L}\left(\left\{(\pi^{k,+} - P\pi^k)\right\}_{k = 1, \dots, K} \right) = \sum_{k = 1}^K \|\pi^{k,+} - P\pi^k\|_2^2
\end{equation}
where $\|\cdot\|_2$ denotes the Euclidean norm. This choice relates to the ``conditional least squares''  method for estimating Markov chain models \cite{hernandez2012adaptive}. Using the loss function \eqref{equ:8}, the optimal solution of $P$ minimizes the average squared error of all data points. If the sum in \eqref{equ:8} is replaced with $\max_{k = 1, \dots, K}$, then the optimal solution minimizes the largest error of all data points. We note that in both cases \eqref{equ:7} can be reformulated as a convex linearly or quadratically constrained quadratic program and be solved using off-the-shelf solvers. In general, the most appropriate choice of the loss function depends on the application of the model.

An alternative method for calibrating the transition matrix~$P$ is ``frequency count,'' which is through the equation:
\begin{equation}\label{equ:9}    
P_{ij} = \frac{S_{ij}}{S_{0j}}
\end{equation}
where $S_{ij} = \sum_{k = 1}^K g_{\text{mc},j}(x^k) g_{\text{mc},i}(x^{k,+})$ and $S_{0j} = \sum_{i = 1}^N S_{ij}$. When indicator functions $\mathbb{I}_{X_i}(x)$ are used as encoding functions $g_{\text{mc},i}(x)$, $i = 1,\dots,N$, $\frac{S_{ij}}{S_{0j}}$ represents the frequency with which the state $x$ transitions to the cell $X_i$ when $x$ is originally in the cell $X_j$. This ``frequency count'' method is introduced in \cite{filev2013generalized} and \cite{li2020fuzzy}.

A calibrated Markov chain model predicts a future probability distribution over $\mathcal{S}$ according to \eqref{equ:4}. A reverse conversion is needed to obtain a deterministic prediction of system state $x$ from the predicted distribution vector $\pi$, called ``decoding.'' The decoding process can be expressed as
\begin{equation}\label{equ:10}    
x = \hat{g}_{\text{mc}}^{-1}(\pi)
\end{equation}
where $\hat{g}_{\text{mc}}^{-1}$ is a decoding function. We use the superscript $-1$ to indicate that this function acts like an inverse of the encoding function $g_{\text{mc}}$ but we add a hat above to signify that it may not be the inverse function of $g_{\text{mc}}$ in the mathematical sense, which typically does not exist due to the dimensional difference between $x$ and~$\pi$. The decoding function may be designed based on the expectation~\cite{filev2013generalized}. For instance, when the state $x$ is encoded to $\pi$ using Gaussian kernel functions as in \eqref{equ:6}, a reasonable design of the decoding function is
\begin{equation}\label{equ:11}    
\hat{g}_{\text{mc}}^{-1}(\pi) = \sum_{i = 1}^N \pi_i \bar{x}_i
\end{equation}
where $\pi_i$ denotes the $i$th entry of $\pi$, and $\bar{x}_i$, $i = 1,\dots,N$, are the centers of the Gaussian distributions. In general, the~pair of encoding and decoding functions, $(g_{\text{mc}}, \hat{g}_{\text{mc}}^{-1})$, should have
\begin{equation}\label{equ:12}    
\varepsilon_{\text{mc}} = x - \hat{g}_{\text{mc}}^{-1}\left(g_{\text{mc}}(x)\right)
\end{equation}
called the ``reconstruction error,'' to be small for all $x$ that are in the range of interest.

\vspace{-1mm}
\subsection{Koopman operator-based modeling}

Assume that state transitions obey the equation
\begin{equation}\label{equ:13} 
x^+ = f(x)
\end{equation}
where $f: \mathbb{R}^n \to \mathbb{R}^n$ is a possibly nonlinear function. Let $\mathcal{G}$ be a linear space of functions $\mathbb{R}^n \to \mathbb{R}$ such that $g \circ f \in \mathcal{G}$ for all $g \in \mathcal{G}$. Note that such a linear space exists -- for instance, the set of all functions $\mathbb{R}^n \to \mathbb{R}$ is such a linear space. Depending on $f$, $\mathcal{G}$ may be finite- or infinite-dimensional~\cite{bruce2019koopman} -- for instance, if $f$ is linear, then $\mathcal{G} = \{x \mapsto c^{\top}x: c \in \mathbb{R}^n\}$ satisfies $g \circ f \in \mathcal{G}$ for all $g \in \mathcal{G}$ and is of dimension $n$. The functions $g$ in $\mathcal{G}$ are called ``observables.'' The Koopman operator, $\mathcal{K}: \mathcal{G} \to \mathcal{G}$, also called the ``composition operator,'' is a linear operator that satisfies
\begin{equation}\label{equ:14} 
\mathcal{K}(g) = g \circ f
\end{equation}
If $\mathcal{G}$ is finite-dimensional of $\text{dim}(\mathcal{G}) = N$ and we let $\mathcal{Z} = \{g_{\text{ko},1},g_{\text{ko},2},\dots,g_{\text{ko},N}\}$ be a basis of $\mathcal{G}$, then the Koopman operator $\mathcal{K}$ can be represented by a matrix $A \in \mathbb{R}^{N \times N}$ that is uniquely determined by
\begin{equation}\label{equ:15} 
A g_{\text{ko}}(x) = g_{\text{ko}}(f(x))
\end{equation}
where $g_{\text{ko}} = [g_{\text{ko},1}, \dots, g_{\text{ko},N}]^{\top}$. Now let 
\begin{equation}\label{equ:16} 
z = g_{\text{ko}}(x)
\end{equation}
which is called the ``lifted state,'' and correspondingly, $z^+ = g_{\text{ko}}(x^+) = g_{\text{ko}}(f(x))$. Then, \eqref{equ:15} leads to
\begin{equation}\label{equ:17} 
z^+ = A z
\end{equation}
which is a linear system propagating the lifted state $z$. We note that $z$ is called the lifted state and the process of \eqref{equ:16} is called ``lifting'' because the dimension of $z$ (which equals the dimension of $\mathcal{G}$ and the number of basis functions in~$\mathcal{Z}$), $N$, is typically higher than that of the original system state~$x$,~$n$.

It is clear that the matrix representation $A$ of the Koopman operator $\mathcal{K}$ in \eqref{equ:15} and \eqref{equ:17} is dependent on the choice of the basis functions $g_{\text{ko},1},\dots,g_{\text{ko},N}$. When the state transition function $f$ in \eqref{equ:13} is unknown, which is typical in a data-driven modeling setting, the observable space $\mathcal{G}$ and its basis~$\mathcal{Z}$ are also not known a priori. A practical strategy is to assume a sufficiently large subspace of $\mathcal{L}^2(\mathbb{R}^n \to \mathbb{R})$\footnote{The Hilbert space of square integrable functions.} to be~$\mathcal{G}$ and choose $N$ elements from a basis for $\mathcal{L}^2(\mathbb{R}^n \to \mathbb{R})$ to be $g_{\text{ko},1},\dots,g_{\text{ko},N}$. This way, as $N$ increases, $\mathcal{Z} = \{g_{\text{ko},1},\dots,g_{\text{ko},N}\}$ increases to a basis of $\mathcal{L}^2(\mathbb{R}^n \to \mathbb{R})$, and the finite-dimensional matrix representation $A$ converges to a Koopman operator $\mathcal{K}$ on $\mathcal{L}^2(\mathbb{R}^n \to \mathbb{R})$. Some possible choices of $g_{\text{ko},1},\dots,g_{\text{ko},N}$ include polynomials up to a certain degree, sinusoidal functions up to a certain frequency, step/indicator functions up to a certain resolution, as well as Gaussian kernel functions \cite{mauroy2019koopman}, given as:
\begin{equation}\label{equ:18} 
g_{\text{ko},i}(x) = \frac{1}{c_i}\, \text{exp} \big(-\frac{1}{2} (x - \bar{x}_i)^{\top} \Sigma^{-1}_i (x - \bar{x}_i)\big)
\end{equation}
where $\bar{x}_i$ and $\Sigma_i$ are the center and the covariance, and $c_i$ is a scaling factor, which are all design parameters.

After the basis functions are chosen, calibrating the Koopman operator model \eqref{equ:17} using system trajectory data is achieved by solving the following optimization problem:
\begin{equation}\label{equ:19}
\min_{A \in \mathbb{R}^{N \times N}}\, \mathcal{L}\left(\left\{(z^{k,+} - Az^k)\right\}_{k = 1, \dots, K} \right)
\end{equation}
where $z^k = g_{\text{ko}}(x^k)$ and $z^{k,+} = g_{\text{ko}}(x^{k,+})$, $k = 1, \dots, K$, are lifted trajectory data, and $\mathcal{L}(\cdot)$ is a loss function that quantifies the error of the Koopman operator model. A~popular choice of the loss function is
\begin{equation}\label{equ:20}    
\mathcal{L}\left(\left\{(z^{k,+} - Az^k)\right\}_{k = 1, \dots, K} \right) = \sum_{k = 1}^K \|z^{k,+} - Az^k\|_2^2
\end{equation}
which minimizes the average squared error of all data points. Using the loss function \eqref{equ:20}, \eqref{equ:19} reduces to a least-squares fitting problem and admits an analytical solution \cite{williams2015data,korda2018linear}. However, it has been shown in~\cite{dahdah2021linear,zhan2022data,kim2023koopman} that alternative choices of the loss function may lead to a model that has a better performance than a least-squares fitting when used for control purposes. In general, the most appropriate choice of the loss function is application-dependent.

A calibrated Koopman operator model predicts a future lifted state $z$ according to \eqref{equ:17}. Similar to Markov chain-based modeling, a ``decoding'' process is needed to obtain a predicted value of $x$ from $z$, which can be expressed as
\begin{equation}\label{equ:21}    
x = \hat{g}_{\text{ko}}^{-1}(z)
\end{equation}
where $\hat{g}_{\text{ko}}^{-1}$ is a decoding function. While a mathematical inverse function of $g_{\text{ko}}$ typically does not exist due to the dimensional difference between $x$ and $z$, a pair of $(g_{\text{ko}}, \hat{g}_{\text{ko}}^{-1})$ should have a small reconstruction error
\begin{equation}\label{equ:22}    
\varepsilon_{\text{ko}} = x - \hat{g}_{\text{ko}}^{-1}\left(g_{\text{ko}}(x)\right)
\end{equation}
for all $x$ that are in the range of interest.

\vspace{-1mm}
\subsection{Computational similarities between the two approaches}

Comparing the Markov chain-based and the Koopman operator-based data-driven modeling approaches introduced above, the following similarities can be observed:
\begin{enumerate}
    \item Both approaches map the original system state $x$ to a new state (a probability distribution vector $\pi$ in Markov chain-based modeling and a lifted state $z$ in Koopman operator-based modeling);
    \item Both approaches predict future values of the new state using a linear system model;
    \item In both approaches the calibration of the linear system model using system trajectory data is achieved by solving an optimization problem;
    \item In both approaches a predicted value of the original state is obtained from a predicted value of the new state through a decoding process.
\end{enumerate}
Furthermore, for mapping the original state $x$ to the new state (called ``encoding'' in Markov chain-based modeling and ``lifting'' in Koopman operator-based modeling), some functions can be used for both approaches, such as indicator/step functions and Gaussian kernel functions, as shown in \eqref{equ:6} and \eqref{equ:18}. Comparing the optimization problems for model calibration of the two approaches, \eqref{equ:7} and \eqref{equ:19}, it can be seen that they are very similar, except that the decision~variable~$P$ in Markov chain-based modeling needs to be a probability transition matrix, i.e., subject to the constraint $P \in \Omega$.

\begin{figure}[h!]
\begin{center}
\begin{picture}(125.0, 60.0)
\put(  0,  -8){\epsfig{file=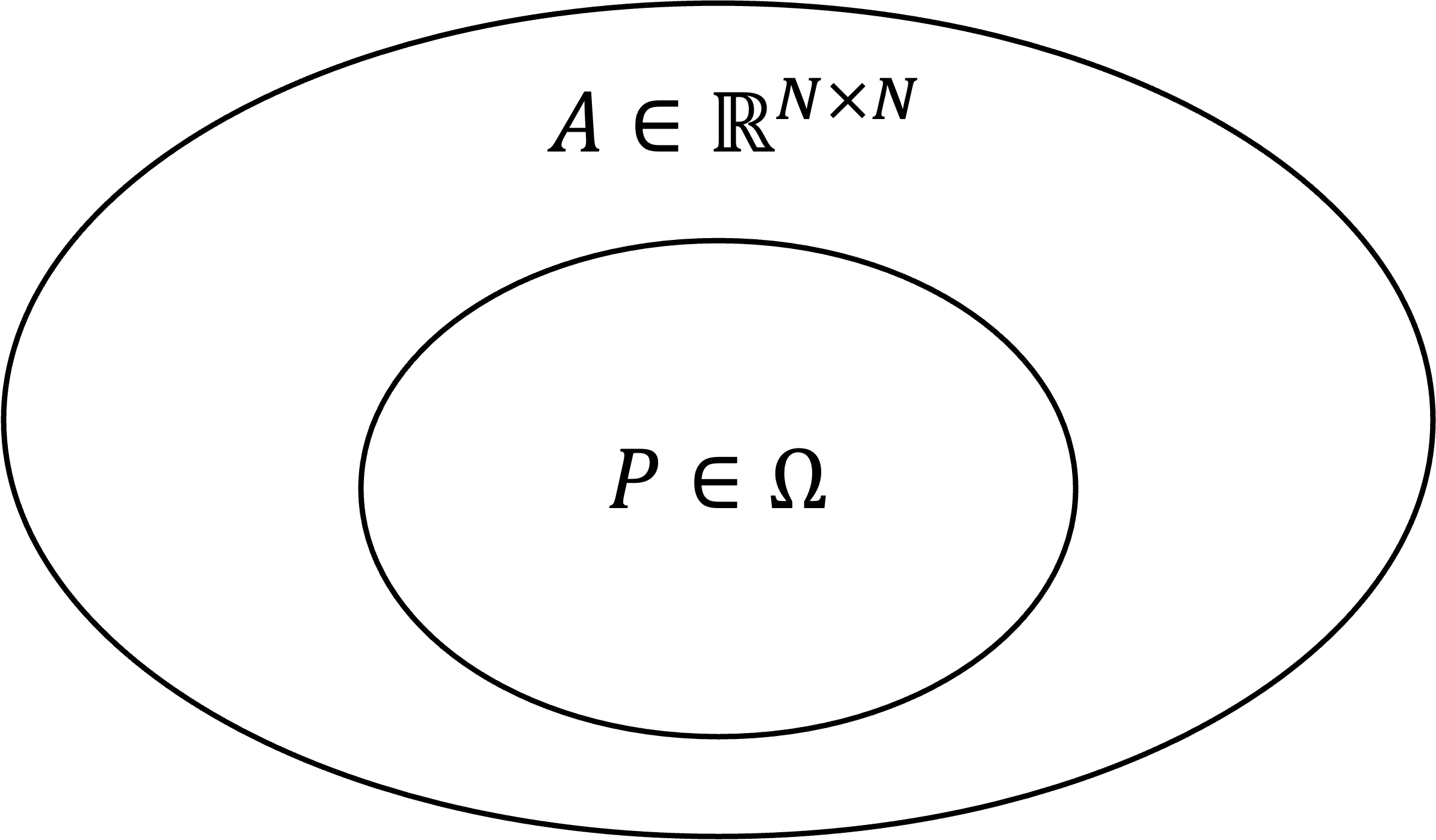,height=.14\textwidth}}
\small
\normalsize
\end{picture}
\end{center}
      \caption{\small{The relation between the spaces for $P$ in Markov chain-based modeling and $A$ in Koopman operator-based modeling.}}
      \label{fig:1}
      \vspace{-2mm}
\end{figure}

{\it Remark~1:} Suppose the same functions (such as Gaussian kernel functions) are used for encoding and lifting (i.e., $g_{\text{mc}} = g_{\text{ko}}$ and, correspondingly, $(\pi^k,\pi^{k,+}) = (z^k,z^{k,+})$) and the same loss function $\mathcal{L}$ is used in the optimization problems \eqref{equ:7} and~\eqref{equ:19}. Then, Koopman operator-based modeling searches over a larger space for an optimal linear model than Markov chain-based modeling, as illustrated in~Fig.~\ref{fig:1}. Consequently, the obtained Koopman operator model will have a greater (or, at least as good) accuracy than the obtained Markov chain model on the data (as measured by the selected loss function). The benefits of a Markov chain model compared to a Koopman operator model include the stochastic process interpretation of the model, together with many useful properties of Markov chains \cite{norris1998markov,meyn2012markov}.

\vspace{-1mm}
\section{Markov chain and Koopman operator-based modeling of controlled systems}

In this section, we discuss Markov chain-based and Koopman operator-based data-driven modeling and corresponding control design methods for controlled systems represented as
\begin{equation}\label{equ:23}
\Sigma: (x,u) \to x^+
\end{equation}
where $x \in \mathbb{R}^n$ and $u \in \mathbb{R}^m$ denote the system state and control input at a certain time instant, $x^+$ denotes the system state at the corresponding next time instant, and $\to$ represents a transition from $x$ to $x^+$ under the control input $u$. Different from the situation for autonomous systems where the two approaches share many similarities, for controlled systems their models and control design methods differ. We will compare them through numerical examples in Section~IV.

\vspace{-1mm}
\subsection{Controlled Markov chain model}

A controlled Markov chain, also called a ``Markov decision process,'' can be understood as a Markov chain the transition probabilities of which depend on an external input, written~as
\begin{equation}\label{equ:24}
p_{ij}(a) \!=\! \mathbb{P}_{a}(s^+ = s_i \,|\, s = s_j) \!=\! \mathbb{P}(s^+ = s_i \,|\, s = s_j, a)
\end{equation}
where the external input $a$ is called an ``action'' and, for computational reasons, is typically assumed to take values in a finite action space $\mathcal{A} = \{a_1,a_2,...,a_M\}$. To represent the original input $u \in \mathbb{R}^m$ using (a combination of) finite actions $a_l \in \mathcal{A}$, similar to state, $u$ is mapped to a distribution on~$\mathcal{A}$ and vice versa through an encoding process $\gamma = h_{\text{mc}}(u)$ and a corresponding decoding process $u = \hat{h}^{-1}_{\text{mc}}(\gamma)$. Using step/indicator functions for encoding corresponds to the conventional discretization strategy based on cell partition. 
 
Using state and action distribution vectors $\pi$ and $\gamma$, a controlled Markov chain model can be compactly expressed~as
\begin{equation}\label{equ:25}
\pi^+ = \sum_{j = 1}^N \sum_{l = 1}^M P_{(\cdot)jl} \pi_j \gamma_l
\end{equation}
where $P$ is now a 3-dimensional tensor and the $(i,j,l)$-entry of $P$ is the transition probability from $s_j$ to $s_i$ under the action $a_l$ (i.e., $P_{ijl} = p_{ij}(a_l)$). Similar to the case of autonomous systems, the calibration of the controlled Markov chain model using system trajectory data can be achieved by solving an optimization problem of the following~form:
\begin{equation}\label{equ:26}
\min_{P \in \Omega}\, \mathcal{L}\Bigg(\bigg\{\Big(\pi^{k,+} - \sum_{j = 1}^N \sum_{l = 1}^M P_{(\cdot)jl} \pi^k_j \gamma^k_l \Big)\bigg\}_{k = 1, \dots, K} \Bigg)
\end{equation}
where $\Omega = \{P \in \mathbb{R}^{N \times N \times M}: P_{ijl} \ge 0, \sum_{i = 1}^N P_{ijl} = 1\}$ is the set of all probability transition tensors, $\pi^k = g_{\text{mc}}(x^k)$, $\gamma^k = h_{\text{mc}}(u^k)$, and $\pi^{k,+} = g_{\text{mc}}(x^{k,+})$, $k = 1, \dots, K$, are encoded trajectory data, and various loss functions $\mathcal{L}$ can be considered for quantifying the model error, such as the average squared error loss function in \eqref{equ:8}.

A significant benefit of modeling a controlled dynamical system as a controlled Markov chain is the availability of various computationally-efficient methods associated with controlled Markov chains for designing controls \cite{puterman2014markov}. We take the value iteration method as an example: Let $c(s,a)$ be a cost function on the Markov state and action spaces $\mathcal{S}$ and~$\mathcal{A}$, which may be designed based on a cost function of the original system state and control input. The value iteration method uses the following equation to estimate an optimal cost-to-go function $V(s)$:
\begin{equation}\label{equ:27}
V^{k+1}(s_j) = \min_{a_l \in \mathcal{A}} \bigg\{ c(s_j,a_l) + \lambda \sum_{i = 1}^N P_{ijl} V^k(s_i) \bigg\}
\end{equation}
where the superscript $k$ indicates the iteration number, and $\lambda \in (0,1)$ is a discount factor. After convergence, an optimal policy $\psi: \mathcal{S} \to \mathcal{A}$ is calculated according to
\begin{equation}\label{equ:28}
\psi(s_j) \in \argmin_{a_l \in \mathcal{A}} \bigg\{ c(s_j,a_l) + \lambda \sum_{i = 1}^N P_{ijl} V^{\infty}(s_i) \bigg\}
\end{equation}
The final step is to decode the policy $\psi$ to a state-feedback controller for the system according to
\begin{equation}\label{equ:29}
u(x) = \hat{h}^{-1}_{\text{mc}}(\gamma_{\psi}(x))
\end{equation}
where $\gamma_{\psi}(x)$ is a distribution on $\mathcal{A}$ dependent on the policy~$\psi$ and the system state $x$, whose $l$th entry, $l = 1,\dots,M$, represents the probability for taking action $a_l$ and is calculated according to $(\gamma_{\psi}(x))_l = \sum_{j:\psi(s_j) = a_l} (g_{\text{mc}}(x))_j$.

\vspace{-1mm}
\subsection{Koopman operator-based models}

Assume that the system dynamics obey the following equation augmented by a trivial equation for control input:
\begin{equation}\label{equ:30}
\begin{bmatrix} x^+ \\ u \end{bmatrix} = \bar{f}(x,u) = \begin{bmatrix} f(x,u) \\ u \end{bmatrix}
\end{equation}
where $x^+ = f(x,u)$ describes the dynamics from $(x,u)$ to~$x^+$, and $\bar{f}$ is $f$ augmented with a projection map from $(x,u)$ to $u$. Consider a Koopman operator $\mathcal{K}: \mathcal{G} \to \mathcal{G}$ for $\bar{f}$,
\begin{equation}\label{equ:31}
\mathcal{K}(g) = g \circ \bar{f}
\end{equation}
where $\mathcal{G}$ is a linear space of functions $\mathbb{R}^{n + m} \to \mathbb{R}$ and $g$ is an observable in $\mathcal{G}$. As before, if $\mathcal{G}$ is of $\text{dim}(\mathcal{G}) = N + M$ and $\mathcal{Z} = \{g_{\text{ko},1},g_{\text{ko},2},\dots,g_{\text{ko},N+M}\}$ is a basis of $\mathcal{G}$, then $\mathcal{K}$ has a matrix representation $\bar{A}$ that is determined by
\begin{equation}\label{equ:32}
\bar{A} g_{\text{ko}}(x,u) = g_{\text{ko}}(\bar{f}(x,u))
\end{equation}
where $g_{\text{ko}} = [g_{\text{ko},1}, \dots, g_{\text{ko},N+M}]^{\top}$. Now we consider two cases, which lead to two different types of models:

\begin{enumerate}[leftmargin=0.46cm]
    \item \textbf{Linear model:} Assume the first $N$ basis functions, $g_{\text{ko},1},$ $\dots,g_{\text{ko},N}$, are nonlinear functions of $x$ and the last $M = m$ basis functions, $g_{\text{ko},N+1},\dots,g_{\text{ko},N+M}$, are projections of $(x,u)$ to the entries of $u$. Then, \eqref{equ:32} can be written as
    \begin{equation}\label{equ:33}
    \begin{bmatrix} z^+ \\ u \end{bmatrix} = \begin{bmatrix} A & B \\ 0 & I \end{bmatrix} \begin{bmatrix} z \\ u \end{bmatrix}
    \end{equation}
    where $z = [g_{\text{ko},1}(x), \dots, g_{\text{ko},N}(x)]^{\top}$, $z^+ = [g_{\text{ko},1}(x^+),$ $\dots, g_{\text{ko},N}(x^+)]^{\top}$, and $A$, $B$, $0$ and $I$ are the $(1,1)$, $(1,2)$, $(2,1)$ and $(2,2)$ blocks of $\bar{A}$. The first $N$ rows of \eqref{equ:33} yields the following linear system of the lifted state $z$:
    \begin{equation}\label{equ:34}
    z^+ = A z + B u
    \end{equation}    
    The calibration of the linear model \eqref{equ:34} using system trajectory data is achieved by solving an optimization problem similar to \eqref{equ:19}:
    \begin{equation}\label{equ:35} \small
    \min_{A \in \mathbb{R}^{N \!\times\! N}, B \in \mathbb{R}^{N \!\times\! m}}\!\! \mathcal{L}\left(\left\{(z^{k,+} - Az^k - Bu^k)\right\}_{k = 1, \dots, K} \right)
    \end{equation} \normalsize

    \item \textbf{Bilinear model:} Assume the first $N$ basis functions are nonlinear functions of $x$, the next $m$ are projections to the entries of $u$, and the last $N m$ are products of each of the first $N$ basis functions and each entry of $u$. Then, \eqref{equ:32} yields the following bilinear system for the lifted state $z = [g_{\text{ko},1}(x), \dots, g_{\text{ko},N}(x)]^{\top}$:
    
    \begin{equation}\label{equ:36}
    z^+ = A z + B u + \sum_{j = 1}^N \sum_{l = 1}^m H_{(\cdot)jl} z_j u_l
    \end{equation} 
    with $H \in \mathbb{R}^{N \times N \times m}$. The calibration of \eqref{equ:36} using system trajectory data is similar to the linear model case:
    \begin{equation}\label{equ:37}
    \min_{A, B, H}\, \mathcal{L}\left(\left\{(z^{k,+} - \hat{z}^{k,+})\right\}_{k = 1, \dots, K} \right)
    \end{equation}
    where $\hat{z}^{k,+} = Az^k + Bu^k + \sum_{j = 1}^N \sum_{l = 1}^m H_{(\cdot)jl} z^k_j u^k_l$.
\end{enumerate}
As before, the loss function $\mathcal{L}$ in \eqref{equ:35} and \eqref{equ:37} for quantifying the model error shall be chosen depending on the application. We note that the assumptions about the forms of the basis functions in the above two cases are restrictive: These basis functions will not increase to a basis of $\mathcal{L}^2(\mathbb{R}^{n + m} \to \mathbb{R})$ as $N$ increases. Consequently, depending on $f$, the error between the actual system dynamics and the models \eqref{equ:34} and~\eqref{equ:36} after calibration may not decrease to zero by using more basis functions in the restricted forms~\cite{bakker2019koopman}. However, considering more general basis functions will make the resulting model still a general nonlinear system of $z$ and~$u$, presenting no computational advantages.

After calibration, the system dynamics are represented by a linear model \eqref{equ:34} or a bilinear model \eqref{equ:36}. Then, methods for linear or bilinear systems can be used for designing controls. A popular method is the MPC \cite{korda2018linear,bruder2020data}. As an example, for regulation problems, control inputs are determined by repeatedly solving the following finite-horizon optimal control problem at each sample time instant~$t$:
\begin{subequations}\label{equ:38}
\begin{align}    
    \min_{u_{\tau}} &\,\, J = \sum_{\tau = 0}^{T-1} \Big(\hat{g}_{\text{ko}}^{-1}(z_{\tau+1})^{\top} Q \hat{g}_{\text{ko}}^{-1}(z_{\tau+1}) + u_{\tau}^{\top} R u_{\tau}\Big) \label{equ:38_1} \\
    \text{s.t.}\, &\,\, z_{\tau+1} = \mathcal{M}(z_{\tau}, u_{\tau}),\,\, \tau = 0, \dots, T-1 \label{equ:38_2} \\
    &\,\, C \hat{g}_{\text{ko}}^{-1}(z_{\tau+1}) \le c,\,\, D u_{\tau} \le d,\,\, \tau = 0, \dots, T-1 \label{equ:38_3} \\
    &\,\, z_0 = [g_{\text{ko},1}(x(t)), \dots, g_{\text{ko},N}(x(t))]^{\top} \label{equ:38_4} 
\end{align}
\end{subequations}
where the cost function in \eqref{equ:38_1} penalizes the errors between the predicted states $x_{\tau+1} = \hat{g}_{\text{ko}}^{-1}(z_{\tau+1})$ and zero and the control effort weighted by the matrices $Q$ and $R$ over the prediction horizon $\tau = 0, \dots, T-1$, the function $\mathcal{M}$ in~\eqref{equ:38_2} represents either the linear model \eqref{equ:34} or the bilinear model~\eqref{equ:36}, the inequalities in \eqref{equ:38_3} represent state/control path constraints (if any), and the initial condition $z_0$ is the lifted state corresponding to measured current system state~$x(t)$. When the decoding function $\hat{g}_{\text{ko}}^{-1}$ is linear in $z$ and $\mathcal{M}$ is the linear model \eqref{equ:34}, the optimization problem~\eqref{equ:38} can be reformulated as a quadratic program (QP) and efficiently solved using off-the-shelf QP solvers. When $\mathcal{M}$ is the bilinear model \eqref{equ:36}, \cite{bruder2021advantages} proposes to fix the value of the lifted state $z$ in the bilinear terms to be $z_0$ to produce a linear approximation of the bilinear model over the prediction horizon. With this approximation, \eqref{equ:38_2} becomes a set of linear equalities, and then \eqref{equ:38} can be solved as~a~QP.

\vspace{-1mm}
\section{Numerical experiments}

We use numerical examples to illustrate and compare the Markov chain-based and the Koopman operator-based data-driven modeling approaches introduced in previous sections. The examples are based on the Van der Pol oscillator system:
\begin{equation}\label{equ:39}
\begin{aligned}
&\dot{x}_1 = x_1 - \frac{1}{3} x^3_1 - x_2 \\
&\dot{x}_2 = x_1 + u
\end{aligned}
\end{equation}
where $x = (x_1, x_2)$ are the system states, and $u$ is an external input. We first use the two approaches to model the free-motion dynamics of the system by fixing $u = 0$, and then use them to model the forced-motion dynamics by incorporating~$u$. For the latter, we also design controls for $u$ based on the obtained models using the methods introduced in Section~III and compare the control performance. In all experiments, we use the continuous-time equations of motion~\eqref{equ:39} to generate trajectory data $(x^k, x^{k,+})$ or $(x^k, u^k, x^{k,+})$, where $x^k$ and $x^{k,+}$ are sampled state values with a sampling period of $0.1$ sec and $u$ is kept constant (either $u = 0$ or $u = u^k$) over the sampling period. 

\vspace{-1mm}
\subsection{Modeling of autonomous systems}

It is known that the unforced Van der Pol oscillator (i.e.,~\eqref{equ:39} with $u = 0$) is a strongly nonlinear system that has an unstable equilibrium at the origin and a globally attractive limit cycle. No linear model of $x$ is able to capture such nonlinear behavior. We use the Markov chain-based and the Koopman operator-based approaches introduced in~Section~II to model the unforced Van der Pol oscillator. To~illustrate the similarities between the two approaches, we choose to use the same set of Gaussian kernel functions to encode/lift the state in both approaches. In particular, we use $81$ Gaussian kernel functions with evenly distributed centers and a common covariance matrix (which has been tuned to achieve the best performance). For model calibration, we use the same set of trajectory data $\{(x^k, x^{k,+})\}_{k = 1,\dots,10^5}$ and the same average squared error loss function $\mathcal{L}$ in the optimization problem in both approaches. We note that in this case the only difference between the optimization problems of the two approaches is that the decision variable $P$ is constrained to~$\Omega$ in Markov chain-based modeling and the decision variable~$A$ can take any value in $\mathbb{R}^{N \times N}$ in Koopman operator-based modeling, as illustrated in Fig.~\ref{fig:1}. Finally, we choose to use the expectation function \eqref{equ:11} for decoding in both approaches. We note that the performance of the calibrated model depends on the choices of the encoding/lifting, decoding, and loss functions in both approaches. Our goal here is not to find the best choices of these functions to produce the most accurate model, but to illustrate the two approaches and their similarities/relations.

\begin{figure}[h!]
\begin{center}
\begin{picture}(180.0, 120.0)
\put(  0,  -10){\epsfig{file=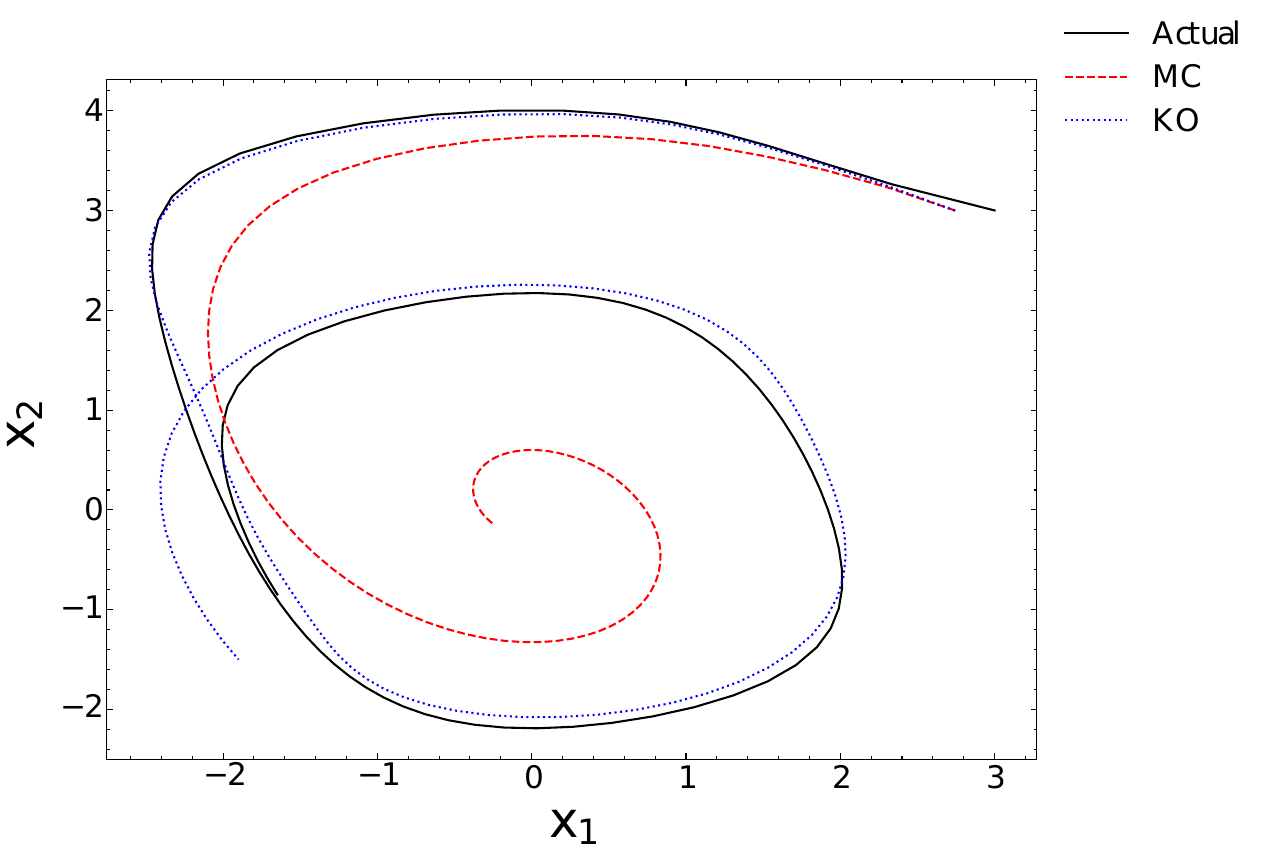,height=.27\textwidth}}
\small.
\normalsize
\end{picture}
\end{center}
      \caption{\small{Actual trajectory of the unforced Van der Pol oscillator system versus predicted trajectories from Markov chain and Koopman operator based models starting from initial condition $x_0 = (3,3)$.}}
      \label{fig:2}
      \vspace{-0.1in}
\end{figure}

After the models are obtained, we simulate the original system, \eqref{equ:39} with $u = 0$, and the models to validate and compare their accuracy. Fig.~\ref{fig:2} shows the actual trajectory (black) versus predicted trajectories from the Markov chain-based model (red) and the Koopman operator-based model (blue) starting from the same initial condition $x_0 = (3,3)$ and over $10$ seconds. It can be seen that the predicted trajectory from the Koopman operator-based model closely matches the actual trajectory over an extended period of time. This close match shows the high accuracy of the model. Meanwhile, the predicted trajectory from the Markov chain-based model matches the actual trajectory at the beginning but deviates from it as time increases. In particular, the predicted trajectory converges toward the origin and does not capture the actual trajectory's behavior of converging to a limit cycle. This convergence toward the origin of the predicted trajectory is largely due to the mixing behavior of a Markov chain model (i.e., the predicted state distribution converges toward a stationary distribution over long run).

In this experiment, the Koopman operator-based model demonstrates better prediction accuracy than the Markov chain-based model, which is consistent with Remark~1: The optimal $A$ matrix of the Koopman operator-based model is searched for over a larger space than the optimal $P$ matrix of the Markov chain-based model and hence is able to achieve a lower loss value, which translates into a higher accuracy. We note that the Markov chain-based model may still be useful for predicting the system behavior over a shorter horizon.

\vspace{-1mm}
\subsection{Modeling and control of controlled systems}

We now use the two approaches to model the Van der Pol oscillator system with control, \eqref{equ:39}. In both approaches, we use the same set of Gaussian kernel functions as in the uncontrolled case to encode/lift the state and the expectation function \eqref{equ:11} for decoding. In the Markov chain-based approach, we use $9$ Gaussian kernel functions with evenly distributed centers and a common covariance matrix to encode the control input. For the Koopman operator-based approach, we derive both a linear model and a bilinear model by choosing basis functions as described in Section~III.B. For the calibration of each model, we use the same set of trajectory data $\{(x^k, u^k, x^{k,+})\}_{k = 1,\dots,10^5}$ and the average squared error loss function $\mathcal{L}$ in the optimization problem.
 
\begin{figure}[h!]
\begin{center}
\begin{picture}(180.0, 120.0)
\put(  0,  -10){\epsfig{file=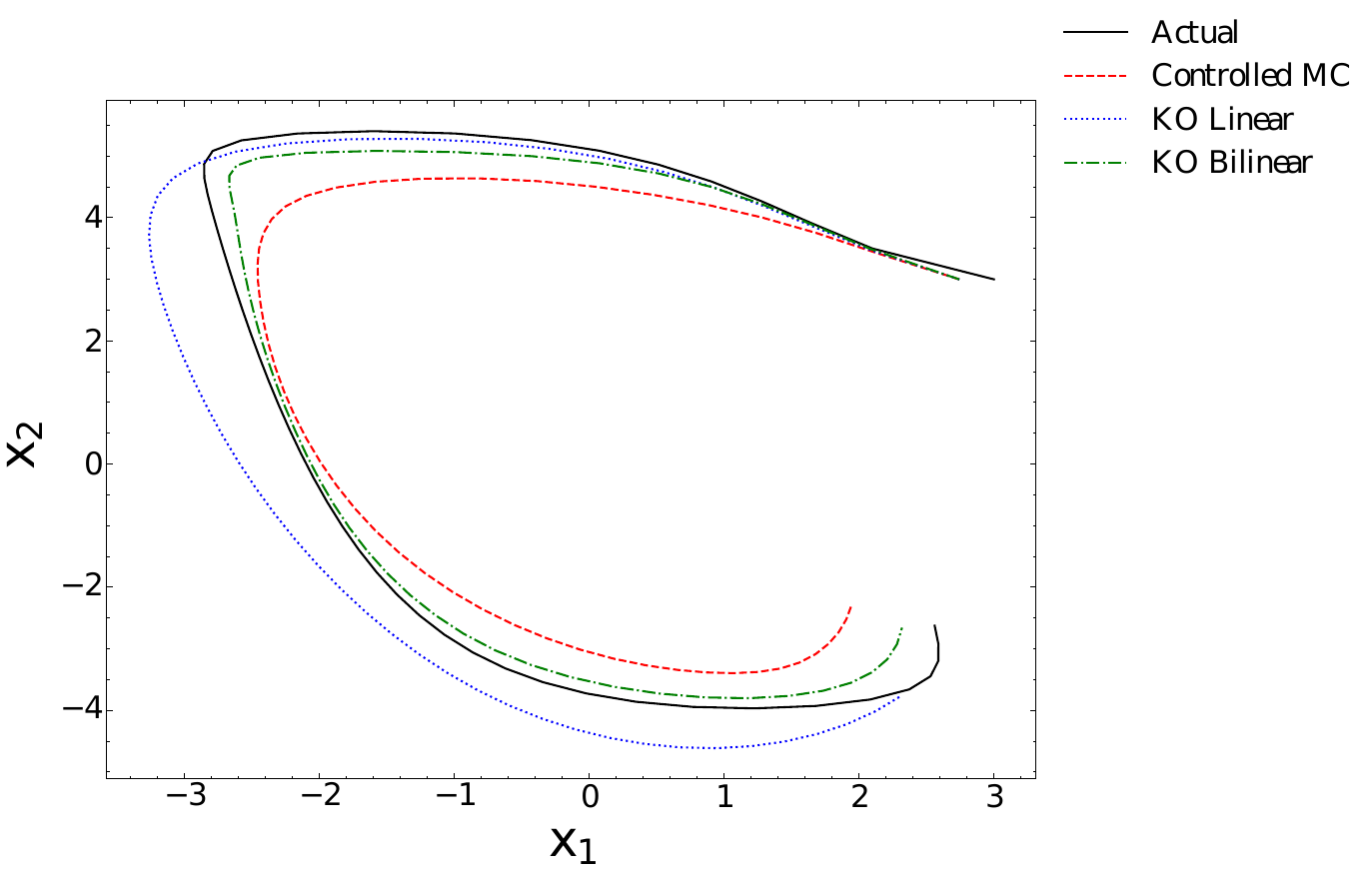,height=.27\textwidth}}
\small
\normalsize
\end{picture}
\end{center}
      \caption{\small{Actual trajectory of the forced Van der Pol oscillator system versus predicted trajectories from controlled Markov chain model and Koopman operator-based linear and bilinear models starting from $x_0 = (3,3)$ and under open-loop input signal $u(t) = 2\cos(t)$.}}
      \label{fig:3}
      \vspace{-0.1in}
\end{figure}

After the models are obtained, we first validate and compare their open-loop prediction accuracy. Fig.~\ref{fig:3} shows the actual trajectory (black) versus predicted trajectories from the controlled Markov chain model (red), the Koopman operator-based linear model (blue), and the Koopman operator-based bilinear model (green) starting from $x_0 = (3,3)$ and under the open-loop input signal $u(t) = 2\cos(t)$ over $5$ seconds. It can be seen that the predicted trajectory from the Koopman operator-based bilinear model matches the actual trajectory most accurately. While the trajectories from the controlled Markov chain model and the Koopman operator-based linear model have larger errors, they are able to match the overall shape and trend of the actual trajectory.

We then design controllers for $u$ based on the obtained models using the methods introduced in Section~III. The goal is to drive the system state $x$ to $0$. For this goal, we consider minimizing a quadratic cost, $c(x,u) = x^{\top} Q x + u^{\top} R u$, with $Q = I$ and $R = 0.5$. For the controlled Markov chain model and the value iteration method, we let $c(s_i,a_l) = c(\bar{x}_i,\bar{u}_l) = \bar{x}_i^{\top} Q \bar{x}_i + \bar{u}_l R \bar{u}_l$, that is, we let the cost associated with each Markov state-action pair $(s_i,a_l)$ be the cost value at their Gaussian kernel centers $(\bar{x}_i,\bar{u}_l)$, and we consider a discount factor of $\lambda = 0.999$. For the Koopman operator-based models and the MPC method, we solve \eqref{equ:38} with a prediction horizon of $T = 50$ and apply the first element, $u(t) = u_0$, over each sampling period. For the bilinear model, we adopt the linear approximation strategy of~\cite{bruder2021advantages} (described below \eqref{equ:38}). Then, for both linear and bilinear models, \eqref{equ:38} is a linear MPC problem and solved as a QP. For comparison, we also implement an MPC based on a nonlinear model of the system obtained from forward Euler discretization of the continuous-time equations of motion~\eqref{equ:39}. This nonlinear MPC uses the same quadratic cost function and horizon length and serves as the ``actual optimal solution.''

\begin{figure}[htbp!]
\begin{center}
\begin{picture}(180.0, 236.0)
\put(  0,  -10){\epsfig{file=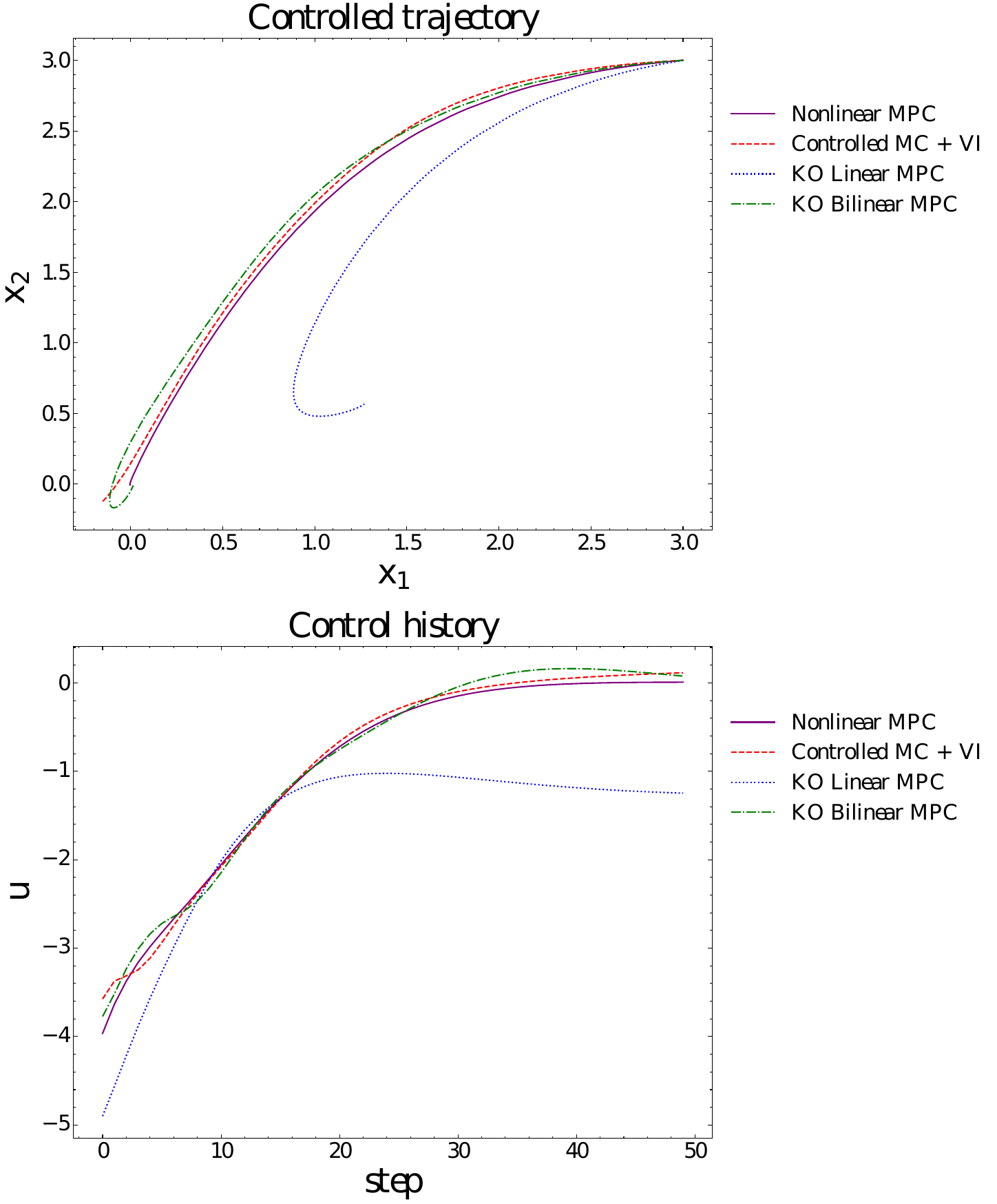,height=.5\textwidth}}
\small
\normalsize
\end{picture}
\end{center}
      \caption{\small{Closed-loop trajectories under the controllers based on controlled Markov chain model + value iteration, Koopman operator-based linear model + linear MPC, Koopman operator-based bilinear model + linear MPC, and nonlinear model + nonlinear MPC.}}
      \label{fig:4}
      \vspace{-0.1in}
\end{figure}

Fig.~\ref{fig:4} shows the closed-loop system state and control input trajectories under the designed controllers starting from the initial condition $x_0 = (3,3)$. It can be seen that the trajectories corresponding to controlled Markov chain model + value iteration and Koopman operator-based bilinear model + linear MPC match the trajectories of nonlinear MPC very well. The state is driven to $0$ (with negligible error) by each of these three controllers. This shows the high effectiveness of the two methods for designing controls. Meanwhile, the controller based on Koopman operator-based linear model + linear MPC fails to drive the state to converge to $0$. This failure is most likely due to notable error between~the Koopman operator-based linear model and the actual nonlinear dynamics near the origin. Such errors are also observed in \cite{bruder2021advantages}, which hence stresses the advantages of bilinear models. In general, the linear model accuracy and the resulting control performance depend on the nonlinear dynamics function $f$ and the choice of the basis functions~$g_{\text{ko}}$. It is shown in \cite{mamakoukas2021derivative} that using higher-order derivatives of $f$ as basis functions can lead to a linear model with bounded errors.
 
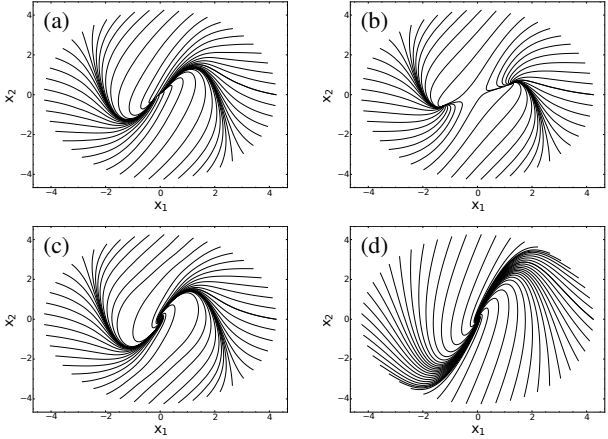
\begin{figure}[h!]
\begin{center}
\begin{picture}(240.0, 155.0)
\put(  0,  75){\epsfig{file=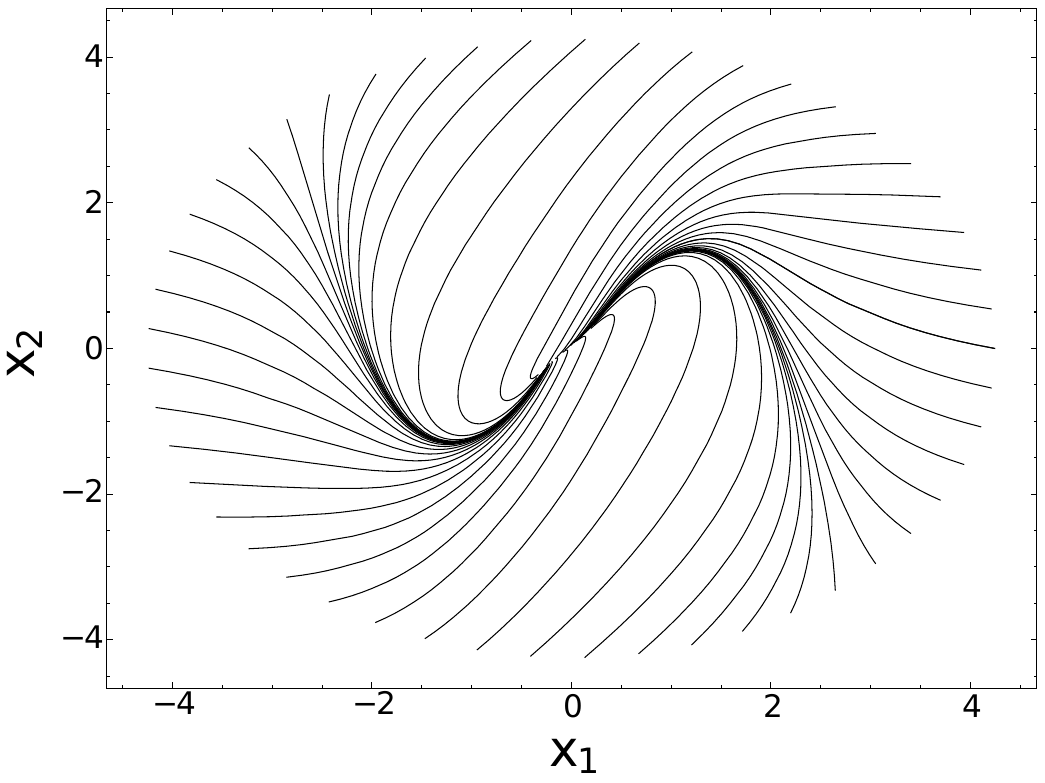,height=.16\textwidth}}
\put(  120,  75){\epsfig{file=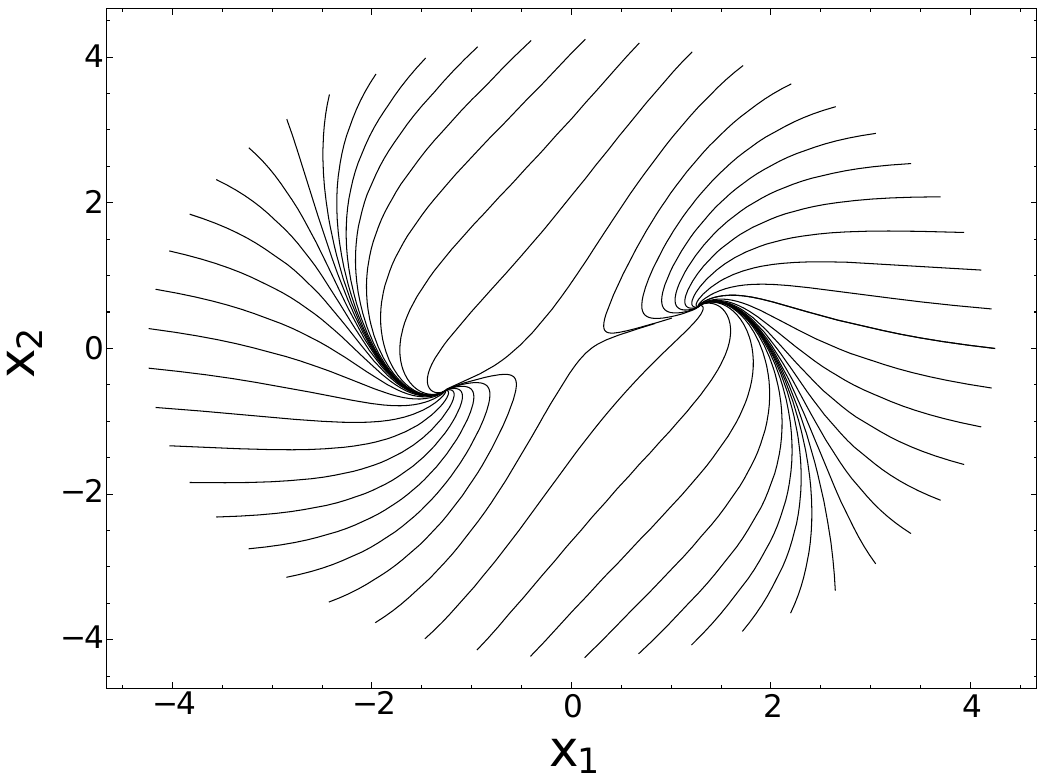,height=.16\textwidth}}
\put(  0,  -10){\epsfig{file=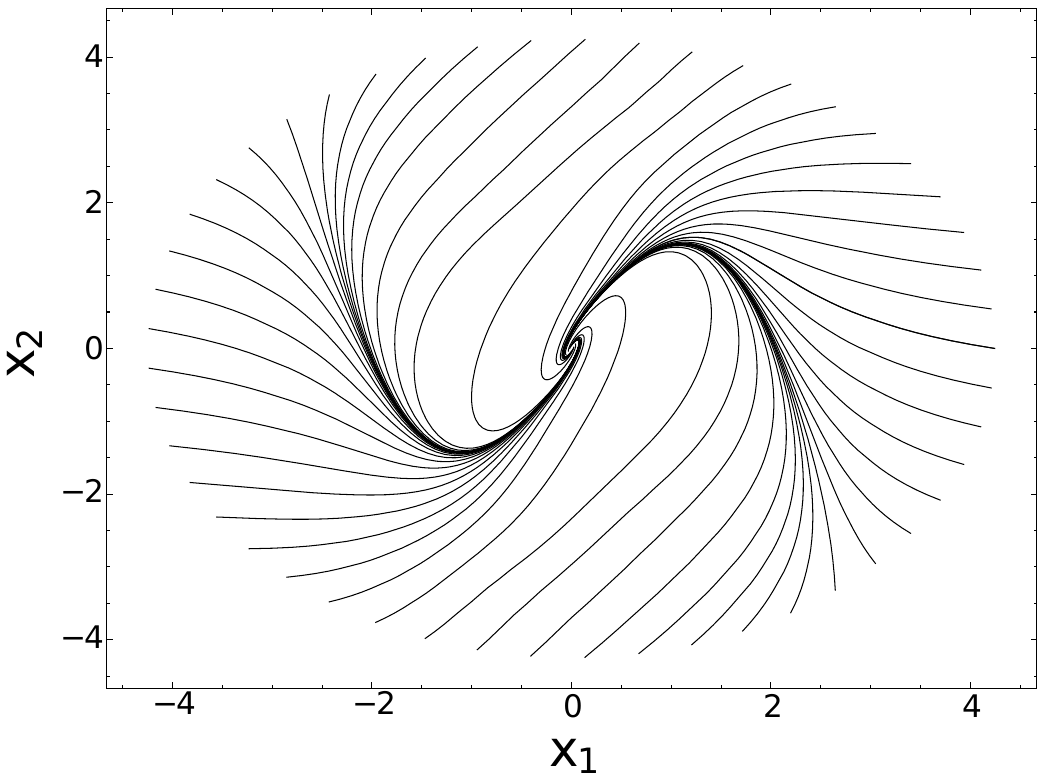,height=.16\textwidth}}
\put(  120,  -10){\epsfig{file=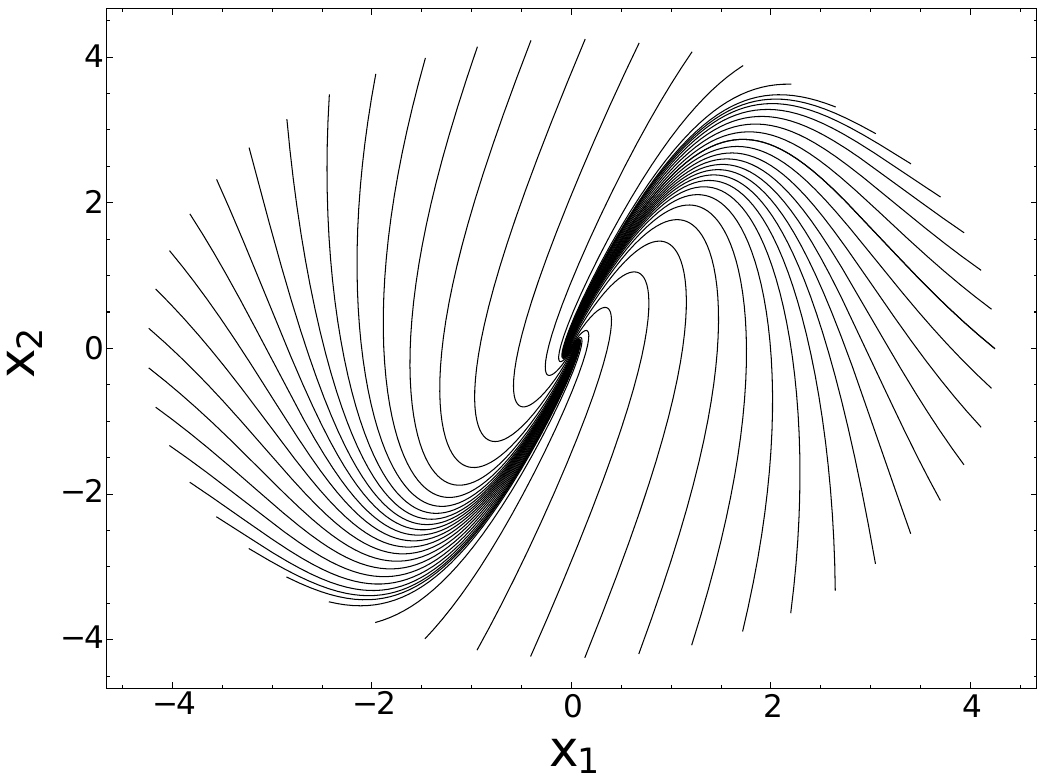,height=.16\textwidth}}

\small
\put(15, 145){(a)}
\put(135, 145){(b)}
\put(15, 60){(c)}
\put(135, 60){(d)}

\normalsize

\end{picture}
\end{center}
\caption{\small{Phase portraits of closed-loop systems under controllers designed from: (a) Controlled Markov chain model + value iteration, (b) Koopman operator-based linear model + linear MPC, (c) Koopman operator-based bilinear model + linear MPC, and (d)~Nonlinear model + nonlinear MPC.}}
      \label{fig:5}
      \vspace{-0.15in}
\end{figure}

To validate the controllers over a larger range of initial conditions, Fig.~\ref{fig:5} plots the phase portrait of the closed-loop system under each controller. The trajectories start from a mesh of initial conditions on a circle of radius $3 \sqrt{2}$. From subplots (a), (c) and (d) it can be seen that, under the controllers designed based on controlled Markov chain model + value iteration, Koopman operator-based bilinear model + linear MPC, and nonlinear MPC, all trajectories converge to the origin, validating these three controllers. From subplot (b) it can be seen that, under the controller based on Koopman operator-based linear model + linear MPC, all trajectories converge to two separate non-zero points. This observation suggests that the closed-loop system under this controller acts like being subject to some constant disturbance that changes its sign depending on the location of the system state. An integral control may be used to cancel such a disturbance, which may be worthy of investigation.

Finally, Table~1 summarizes the computation time of each controller design method. All computations are performed in {\sf Python} on a laptop with Intel Core i7-7700HQ CPU and 16 GB RAM. For the controlled Markov chain model + value iteration method, the time is for value iteration until convergence and producing a feedback control law $u(x)$. Once $u(x)$ has been obtained, it can be directly used to control the system without any further computations except for a simple function evaluation. This fact represents a major computational advantage of the controlled Markov chain model + value iteration method as compared to MPC-based methods. For the other three MPC-based methods, the times are the average time per step for solving the online optimization problem and obtaining a control input value~$u(t)$. Among the three methods, the Koopman operator-based linear model + linear MPC method is the fastest, the Koopman operator-based bilinear model + linear MPC method is the second, and nonlinear MPC is the slowest. The reduced computation time as compared to nonlinear MPC is a significant benefit of the Koopman operator-based methods. The use of the Koopman operator-based bilinear model entails a longer computation time than the use of the Koopman operator-based linear model because the bilinear model involves more basis functions which results in a higher-dimensional model.

\vspace{-2mm}
\begin{table}[h!]
\centering
\caption{Computation times of different control methods.}
\begin{tabular}{|c|c|}
\hline
Method  & Time [ms] \\ \hline
Controlled MC + VI   & 95.4 (total) \\ \hline
KO Linear MPC   & 2.1 (per step) \\ \hline
KO Bilinear MPC & 34.7 (per step) \\ \hline
Nonlinear MPC        & 272.7 (per step) \\ \hline
\end{tabular}
\vspace{-2mm}
\end{table}

\vspace{-1mm}
\section{Summary}

In this paper, we first introduced the computational similarities between Markov chain and Koopman operator-based modeling of autonomous systems. We showed that both approaches achieved linear-like dynamic flow predictions via a similar procedure consisting of an encoding/lifting process, an optimization-based calibration, and an decoding process. We then introduced the models and corresponding control design methods of the two approaches for controlled systems. We used numerical examples to illustrate and compare the two approaches in prediction accuracy and computation efficiency for both autonomous and controlled systems.

\vspace{-0.1mm}
\bibliographystyle{ieeetr}
\bibliography{ref}

\end{document}